# Citation count distributions for large monodisciplinary journals[1]

*Mike Thelwall*, Statistical Cybermetrics Research Group, University of Wolverhampton, UK.

Many different citation-based indicators are used by researchers and research evaluators to help evaluate the impact of scholarly outputs. Although the appropriateness of individual citation indicators depends in part on the statistical properties of citation counts, there is no universally agreed best-fitting statistical distribution against which to check them. The two current leading candidates are the discretised lognormal and the hooked or shifted power law. These have been mainly tested on sets of articles from a single field and year but these collections can include multiple specialisms that might dilute their properties. This article fits statistical distributions to 50 large subject-specific journals in the belief that individual journals can be purer than subject categories and may therefore give clearer findings. The results show that in most cases the discretised lognormal fits significantly better than the hooked power law, reversing previous findings for entire subcategories. This suggests that the discretised lognormal is the more appropriate distribution for modelling pure citation data. Thus future analytical investigations of the properties of citation indicators can use the lognormal distribution to analyse their basic properties. This article also includes improved software for fitting the hooked power law.

## 1. Introduction

Journals, authors, departments and universities are sometimes evaluated with the aid of indicators derived from citation counts, such as the Journal Impact Factor (JIF) (Garfield, 2006), the h-index (Hirsch, 2005) or the Mean Normalized Citation Score (MNCS) (Waltman, van Eck, van Leeuwen, Visser, & van Raan, 2011). The appropriateness of any indicator depends upon the properties of the data on which it is based (Wang, Song, & Barabási, 2013). For example, the JIF is imprecise because sets of citation counts are highly skewed and its calculation uses the arithmetic mean, which is inappropriate for skewed data sets - the geometric is a better option (Thelwall & Fairclough, 2015; Zitt, 2012).

Knowledge about the statistical distribution that best fits citation data can also aid theoretical understanding of how citations accrue in order to give context to interpretations of scores. This is important because the straightforward explanation that citations reflect relevant contributions from prior work (Merton, 1973) is not the full truth. Citations are affected by factors that are apparently unrelated to the quality of the cited work, such as the number of prior citations (Merton, 1968) as well as the nationality of the authors in collaborations (Glänzel, 2001), and document-based properties, such as the readability of the abstract (Gazni, 2011). Identifying the influence of such factors requires, at least in part, a statistical approach in order to detect tendencies that may not be evident in individual articles (Didegah & Thelwall, 2013; Onodera & Yoshikane, 2015). For this, identifying the most appropriate statistical distribution is essential because analyses that use incorrect distributions can reach unjustified conclusions (Thelwall & Wilson, 2014b; Thelwall, 2016a).

It is impossible to logically or empirically *prove* that any given statistical distribution fits citation counts perfectly, which is a generic issue with mathematical models of real data (e.g., Burnham & Anderson, 2002, p. 20). Nevertheless, researchers can assess whether a distribution fits citation counts reasonably well and can also compare two or more distributions to check which fits best. Although some such attempts exclude articles with few citations and have found that the remaining articles fit single parameter distributions well, such as the power law and the Yule-Simon process (Brzezinski, 2015; Clauset, Shalizi, & Newman, 2009), this does not help citation analysis in practice because uncited and low-cited articles are rarely completely ignored by citation-based indicators (the h-index is an exception). When including low cited articles and uncited articles, the shifted/hooked power law (for background see: Pennock, Flake, Lawrence, Glover, & Giles, 2002) and discretised lognormal distributions (for continuous lognormal background see: Limpert, Stahel, & Abbt, 2001) fit substantially better (Eom & Fortunato, 2011; Evans, Kaube, & Hopkins, 2012; Radicchi, Fortunato, & Castellano, 2008; Thelwall & Wilson, 2014a; Thelwall, 2016a) and the lognormal distribution seems to have the wrong shape for subject categories (Thelwall, 2016b). The negative binomial distribution has also been suggested but does not fit as well as the hooked power law and discretised lognormal distributions (Low, Thelwall, & Wilson, 2015). Stopped sum models have been found to fit better on some data sets but have parameter estimation problems (Low, Thelwall, & Wilson, 2015), as does the hooked power law in a minority of cases. Models have also been proposed for predicting the growth of citations over time (Yao, Peng, Zhang, & Xu, 2014; Wu, Fu, & Chiu, 2014), with one suggesting that the lognormal may not be appropriate for individual articles with a long term total citation count above 8.5 (Wang, Song, & Barabási, 2013).

[1] Thelwall, M. (in press). Citation count distributions for large monodisciplinary journals. Journal of Informetrics.

Although the hooked power law and discretised lognormal distribution seem to be the best distributions found so far for citation analysis, in terms of their fit to citation data and (relative) robustness of parameter estimation, studies so far suggest that each one is preferable to the other in some subject areas but not in others. If uncited articles are excluded, then the hooked power law fits better than the discretised lognormal for 15 of out 20 varied Scopus categories for journal articles from 2004 (Thelwall & Wilson, 2014a). If no articles are excluded then a similar conclusion holds: the hooked power law is a better fit than the discretised lognormal for 22 out of 26 varied Scopus categories for journal articles from 2009, although the discretised lognormal fits better than the hooked power law for a larger percentage of categories for more recent articles (Thelwall, 2016a). The hooked power law has been found to fit better than the discretised lognormal for a set of ten physics journals, using different subsets of articles from 1950 to 2008 (Eom & Fortunato, 2011). The (not discretised) lognormal has also been shown to fit articles from 20 different Web of Science subject categories reasonably well (Radicchi, Fortunato, & Castellano, 2008). A limitation of the first two studies is that Scopus subject categories can include journals with very different specialisms within a field and if any of these specialisms have different citation properties then the overall subject category citation distribution will be impure. The third study investigated only one subject area, physics, and the fourth did not compare the hooked power law with the lognormal distribution.

A logical way around the problem of impure subject categories is to select single journals rather than entire subject categories. Non-general journals often target a specific field and hence should have a narrower focus than collections of journals within a subject. Some studies have adopted this strategy (e.g., using *Physical Review D:* Redner, 1998), but none have compared the discretised lognormal with the hooked power law without excluding low cited articles. Moreover, larger scale systematic studies across disciplines (e.g., not restricted to physics) are needed to make general conclusions possible. This study fills this gap by analysing a set of 50 different large non-general journals to see whether there is evidence that one of the two models tends to fit these purer distributions better than the other. This would give evidence that the better fitting distribution is the pure distribution whereas the other may only fit subject categories that are impure. Large amounts of data are needed to get accurate fits of statistical models and so the 50 non-general journals with the most articles indexed in Scopus were selected. The research question is therefore the following.

- RQ: Which out of the hooked power law and the discretised lognormal distribution is the best fitting for sets of citation counts from articles published in large non-general journals?

This article uses a similar main strategy to a previous paper (Thelwall, 2016b) but uses a new and different type of data set (journals rather than subject categories), has an improved method for fitting the hooked power law, and reaches different conclusions.

## 2. Methods

**Data**: To identify the journals with the most articles in Scopus, the query `PUBYEAR IS 2006 AND DOCTYPE(ar)` was run to match all journal articles from 2006. The year 2006 was chosen to give a decade to attract citations so that the citation distribution should be mature and there should not be a substantial difference between articles published early in the year compared with articles published late in the year. Scopus was selected in preference to the Web of Science for its larger coverage of academic literature (Li, Burnham, Lemley, & Britton, 2010; López-Illescas, de Moya-Anegón & Moed, 2008; Moed & Visser, 2008). The Refine option was then used to identify the 50 titles with the most matching articles. One conference proceedings (IEEE Engineering in Medicine and Biology Society Conference) was rejected and replaced with the next largest journal. Although two publications were magazines rather than traditional academic journals (Jane's magazines) they were not classified as such in Scopus and were retained in order to include a contrasting type of publication. None had to be excluded for being generalist (e.g., Nature, Science or PLoS ONE). A complete list of journals and their Scopus categories is given in Appendix B. The citation counts for all records of type article were extracted for each journal for 2006 from Scopus on 27-29 January 2016.

**Distribution fitting**: The probability mass function of the hooked power law is:

$$h(n) = (B+n)^{-\alpha} / \sum_{n=1}^{\infty} (B+n)^{-\alpha}$$

with free parameters α and $B$. The hooked power law is a discrete version of the Lomax distribution (Lomax, 1954), which is a special case of the Pareto type II distribution (Burrell, 2008) and so it is reasonably well-known in its continuous form.

The probability density function of the continuous lognormal distribution is:

$$c(x) = \frac{1}{x\sigma\sqrt{2\pi}} e^{-\frac{(\ln(x)-\mu)^2}{2\sigma^2}}$$

with scale parameter μ and location parameter σ. These are the mean and standard deviation of the natural log of the data (Limpert, Stahel, & Abbt, 2001). The probability mass function of the discretised lognormal distribution is:

$$d(n) = \int_{n-0.5}^{n+0.5} c(x)dx \,/ \int_{0.5}^{\infty} c(x)dx$$

Before fitting any distributions, 1 was added to all citation counts so that the discretised lognormal distribution could be fitted to the complete data sets. It makes no difference to the hooked power law, except for altering the value of the B parameter by 1. The discretised lognormal was fitted using the R powerRlaw package (Gillespie, 2015) and the hooked power law was fitted using an extended version of previously-written R code (Thelwall & Wilson, 2014a) – see below for details of the extension. The fit of the two distributions was compared using the log-likelihood, with the highest value indicating the better fit. This is equivalent to the standard AIC test (Akaike, 1974) because the models have the same number of free parameters (two each). With the log-likelihood approach, the best model is the one that gives the highest probability of generating the dataset. The statistical significance of the difference in fits between the two models was assessed with the Vuong test (Vuong, 1989), which also uses the log-likelihood approach and is appropriate because the distributions are non-nested. An alternative way of testing the fit of a distribution to a dataset is to compare the empirical cumulative distribution function with the theoretical cumulative distribution function of the model.

Fitted distributions were compared to the empirical data through plots of their cumulative distributions on a common graph. Although visual examinations of Q-Q plots are the standard technique for such comparisons, cumulative distribution functions have the advantage for discrete data that important values (such as 1) are clearly distinguished. They are also easier for non-experts to interpret. Visual comparisons are useful to check for unusual patterns in model fits. This approach allows systematic differences in shape between the empirical and model distributions to be identified. The discovery of any such patterns would be stronger evidence than a goodness of fit test that a distribution did not have the correct *shape* to fully describe a data source. This is because systematic differences in shape suggest that the problem is not just size of random factors within the data but that the process producing the data has inherent properties that the distribution cannot model. Since there are many fits to check (100), following an initial visual check (not shown) the following heuristic was generated to automatically check them in a way that simulated, but made systematic, visual comparisons.

In order to automatically test for differences between the empirical and theoretical cumulative distributions, four non-overlapping intervals were generated and the maximum magnitude of the difference between the empirical and theoretical distribution calculated within each interval. Four intervals were chosen because the initial visual examination suggested that this number would illustrate systematic differences in shape most clearly. Exact quantiles cannot be calculated for discrete data and so an alternative method was devised to generate four segments. For this method, the intervals were calculated to be approximately equally spaced on a logarithmic scale. For each journal, the interval starting point was set at 1 (i.e., $ln(1+0)$ for uncited articles) and the ending point was set at $ln(1+n_{max})$, where $n_{max}$ is the maximum citation count for the journal. The intervals were then set to be equally spaced on the logarithmic scale, with the starting point rounded up and the ending point rounded down to avoid overlaps between intervals. The equal spacing was assigned on the logarithmic scale rather than for the original citation counts because this better captures the key shape changes in the distributions.

See Appendix A for steps taken to maximise the accuracy of the parameters when fitting the hooked power law distribution.

## 3. Results

For the complete data set, not all of the hooked power law fits converged (Table 1), with 9 out of 50 reaching the 10k limit for α set in the system to constrain the calculation time (see

Appendix A). All except one of these had large negative Vuong z scores, suggesting that even after convergence they would be likely to fit the lognormal better than the hooked power law. The remaining journal, Jane's Defence Industry, may change to give a significant Vuong test for the hooked power law after convergence. To check that the 10k α limit was unlikely to alter the Vuong test result in most cases, the model fitting was repeated with an α limit 100 times smaller, at 10. The difference in log-likelihoods in all except two cases were less than 20% of the difference between the log-likelihoods of the hooked power law and discretised lognormal distributions (see the online supplement cited in Appendix B). This suggests that even enormously larger α values would not eliminate, or even substantially reduce, the difference between the hooked power law and discretised lognormal log-likelihood values. The two exceptions are the two magazines in the collection. Thus, the 10k α limit probably did not influence the overall results, except perhaps for the two magazines.

Overall the discretised lognormal fits the data statistically better than the hooked power law for most (36) journals and the reverse is not true for any journals in the data set because the difference in fits are not statistically significant for the remaining 14 journals. This is strong evidence that the discretised lognormal is a better fit for relatively pure collections of articles.

**Table 1**. Hooked power law and lognormal distributions fitted to counts of citations to articles from 2006 in the selected 50 large journals.

| Journal | Art. | Ln µ | Ln σ | Ln LL | Hk α | Hk B | Hk LL | Vuong | Best |
|---|---|---|---|---|---|---|---|---|---|
| Acta Crystallographica Section E | 4218 | 0.93 | 0.86 | -9159.8 | 11.7 | 30.8 | -9179.4 | -2.58 | L* |
| Angewandte Chemie | 1362 | 3.89 | 0.91 | -7095.7 | 24.2 | 1607.4 | -7193.7 | -7.76 | L* |
| Applied Mathematics & Computation | 1243 | 2.09 | 1.12 | -4490.0 | 6.0 | 54.9 | -4481.9 | 1.52 | H |
| Applied Physics Letters | 6103 | 2.94 | 1.03 | -26825.0 | 7.7 | 175.4 | -26963.0 | -6.59 | L* |
| Applied Surface Science | 1545 | 2.37 | 1.03 | -5899.3 | 7.1 | 88.5 | -5925.7 | -3.04 | L* |
| Astronomy & Astrophysics | 1864 | 2.97 | 1.00 | -8192.4 | 10.5 | 261.1 | -8245.5 | -4.33 | L* |
| Astrophysical Journal | 2688 | 3.36 | 0.97 | -12755.5 | 10.9 | 394.7 | -12878.9 | -7.13 | L* |
| Biochemical & Biophysical Res. Comm. | 2335 | 2.84 | 0.88 | -9650.8 | 11.9 | 245.8 | -9822.4 | -10.00 | L* |
| Biochemistry | 1599 | 3.07 | 0.74 | -6695.0 | 358.2 | 9986.7 | -6929.7 | -12.64 | L* |
| Bioorganic & Medicinal Chemistry Lett. | 1240 | 2.93 | 0.73 | -5007.9 | 10k | 250153 | -5167.0 | -10.12 | L* |
| Brain Research | 1375 | 2.86 | 0.89 | -5725.5 | 22.4 | 511.9 | -5808.7 | -6.56 | L* |
| Cancer Research | 1428 | 4.00 | 0.79 | -7399.7 | 29.5 | 2064.5 | -7593.2 | -11.00 | L* |
| Chemical Physics Letters | 1650 | 2.52 | 0.96 | -6446.7 | 10.8 | 165.7 | -6494.5 | -4.13 | L* |
| Chinese J. of Clinical Rehabilitation | 2668 | -0.30 | 0.77 | -2203.3 | 7.1 | 3.3 | -2203.7 | -0.61 | L |
| Geophysical Research Letters | 1636 | 3.01 | 0.97 | -7193.2 | 9.2 | 225.1 | -7256.1 | -4.68 | L* |
| Inorganic Chemistry | 1432 | 3.25 | 0.85 | -6457.1 | 10k | 363813 | -6567.0 | -7.09 | L* |
| Jane's Defence Industry | 1320 | -0.09 | 0.19 | -38.3 | 10k | 1853.0 | -38.4 | 0.00 | L |
| Jane's Defence Weekly | 1975 | -7.23 | 1.34 | -134.3 | 6.5 | 0.0 | -134.3 | -0.02 | L |
| Japanese J. of Applied Physics Part 1 | 2229 | 1.75 | 1.10 | -7241.5 | 4.6 | 25.6 | -7241.9 | -0.10 | L |
| Jisuanji Gongcheng Computer Eng. | 1945 | -0.35 | 0.97 | -2188.0 | 4.6 | 2.3 | -2188.7 | -0.54 | L |
| J.of Agricultural & Food Chemistry | 1448 | 3.23 | 0.83 | -6465.7 | 108.3 | 3708.5 | -6591.9 | -7.64 | L* |
| J. of Applied Physics | 3570 | 2.32 | 1.11 | -13714.1 | 5.6 | 63.7 | -13712.3 | 0.15 | H |
| J. of Applied Polymer Science | 2438 | 2.24 | 0.95 | -8805.0 | 17.4 | 212.3 | -8838.7 | -2.38 | L* |
| J. of Biological Chemistry | 4306 | 3.62 | 0.75 | -20432.3 | 10k | 492666 | -21050.1 | -21.15 | L* |
| J. of Chemical Physics | 2870 | 2.70 | 1.00 | -11818.1 | 7.1 | 120.1 | -11900.3 | -5.23 | L* |
| J. of Immunology | 1806 | 3.64 | 0.82 | -8782.4 | 155.4 | 8030.2 | -8953.9 | -7.33 | L* |
| J. of Neuroscience | 1325 | 4.12 | 0.73 | -6929.2 | 10k | 816584 | -7144.0 | -14.35 | L* |
| J. of Organic Chemistry | 1469 | 3.25 | 0.80 | -6525.7 | 10k | 348732 | -6666.4 | -8.15 | L* |
| J. of Physical Chemistry A | 1686 | 2.83 | 0.93 | -7040.4 | 11.7 | 244.1 | -7121.1 | -5.61 | L* |
| J. of Physical Chemistry B | 3617 | 3.24 | 1.00 | -16846.7 | 6.8 | 195.6 | -17007.1 | -8.76 | L* |
| J. of Power Sources | 1475 | 3.32 | 0.99 | -6982.0 | 22.9 | 884.8 | -6998.6 | -1.17 | L |
| J. of the American Chemical Soc. | 3254 | 3.99 | 0.88 | -17173.8 | 14.4 | 988.4 | -17483.0 | -15.29 | L* |
| J. of Virology | 1232 | 3.56 | 0.79 | -5855.1 | 10k | 482021 | -5992.2 | -10.87 | L* |
| Langmuir | 1696 | 3.32 | 0.93 | -7898.3 | 13.3 | 466.6 | -8006.1 | -8.88 | L* |
| Macromolecules | 1263 | 3.37 | 0.90 | -5914.2 | 67.1 | 2743.6 | -5988.6 | -5.80 | L* |
| Materials Science & Eng. A | 1490 | 2.68 | 1.00 | -6111.8 | 13.7 | 263.4 | -6133.8 | -2.06 | L* |
| Monthly Not. R. Astronomical Soc. | 1352 | 3.15 | 1.07 | -6280.5 | 5.2 | 131.7 | -6317.9 | -4.06 | L* |
| Nuclear Instruments & Meth. Physics A | 1569 | 1.74 | 1.12 | -5109.7 | 4.2 | 21.6 | -5113.6 | -1.09 | L |
| Optics Express | 1324 | 3.08 | 1.07 | -6045.0 | 8.2 | 221.6 | -6056.6 | -1.21 | L |
| Organic Letters | 1524 | 3.44 | 0.80 | -7068.0 | 10k | 429455 | -7237.2 | -9.57 | L* |
| Physica B Condensed Matter | 1275 | 1.30 | 1.15 | -3622.7 | 3.9 | 12.6 | -3620.9 | 1.02 | H |
| Physical Review A | 2080 | 2.60 | 0.99 | -8330.2 | 22.1 | 409.0 | -8347.5 | -1.30 | L |
| Physical Review B | 5603 | 2.73 | 1.02 | -23358.8 | 5.9 | 98.1 | -23537.3 | -9.70 | L* |
| Physical Review D | 2305 | 2.85 | 1.16 | -10185.3 | 5.4 | 106.1 | -10177.5 | 0.83 | H |
| Physical Review E | 2448 | 2.48 | 1.06 | -9686.4 | 5.7 | 73.5 | -9721.2 | -3.29 | L* |
| Physical Review Letters | 3760 | 3.52 | 0.99 | -18515.2 | 7.7 | 306.1 | -18700.6 | -9.88 | L* |
| PNAS | 3297 | 4.24 | 0.89 | -18292.2 | 18.5 | 1659.6 | -18487.1 | -4.66 | L* |
| Tetrahedron | 1275 | 2.88 | 0.76 | -5139.1 | 64.1 | 1449.4 | -5292.0 | -8.40 | L* |
| Tetrahedron Letters | 1987 | 2.79 | 0.81 | -7951.2 | 10k | 219180 | -8122.1 | -10.96 | L* |
| Thin Solid Films | 1253 | 2.47 | 1.03 | -4907.4 | 9.1 | 133.4 | -4916.4 | -1.02 | L |

The shape of the fitted discretised lognormal distribution seems very close to the shape of the empirical data. Nevertheless, there are some patterns in the way in which the shapes differ (Table 2). In particular, the model tends to slightly overestimate the number of citations in the second smallest group (relatively low cited articles – see Figure 1 for an example) since the mean difference is -1.8% overall for this set. Although the differences are relatively small, the existence of a pattern in the way in which the theoretical distributions tend to not match the empirical data suggests that the discretised lognormal distribution is not quite the right shape for this type of data.

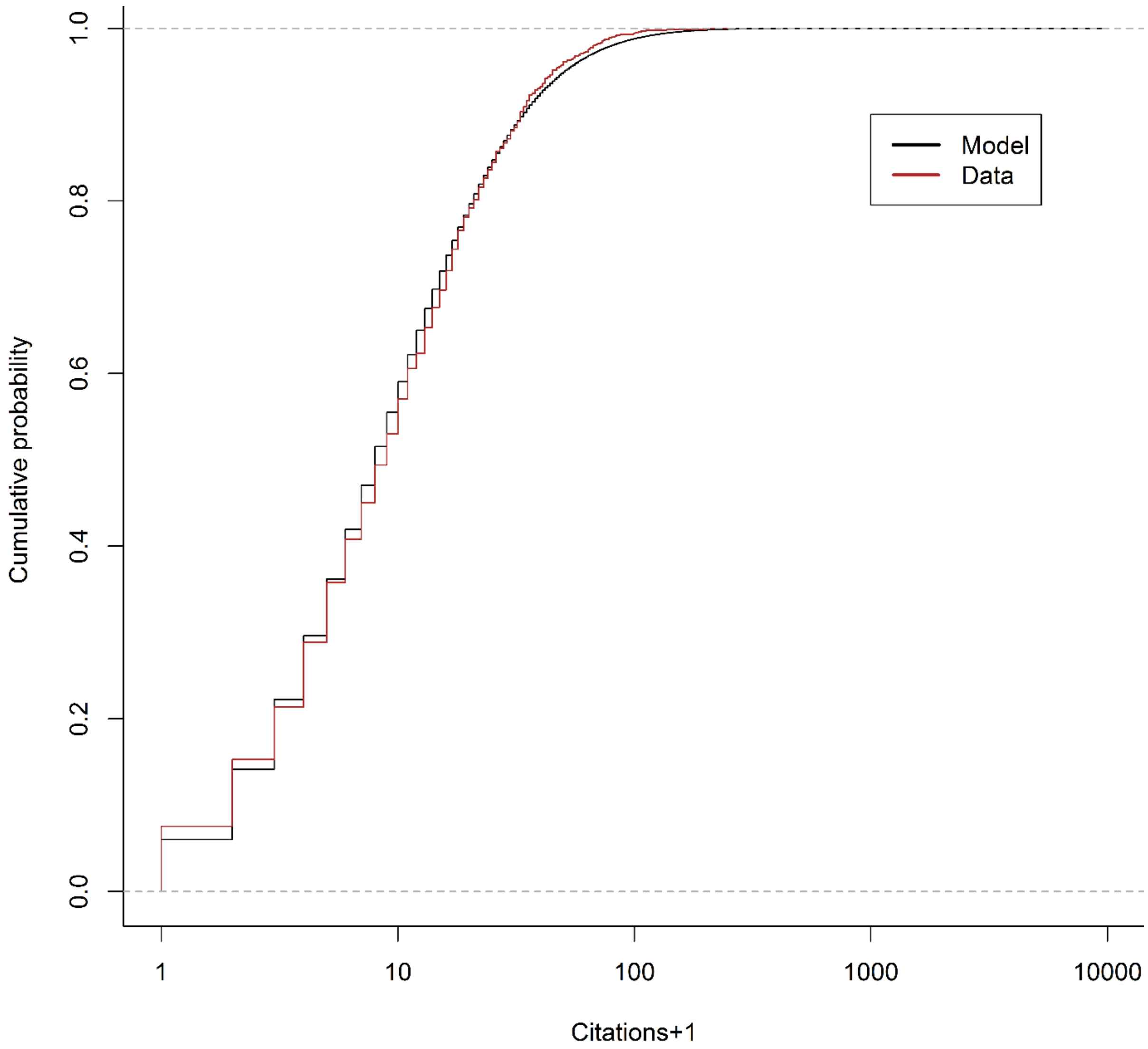

**Figure 1**. The empirical and discretised lognormal model cumulative distribution functions for the journal Applied Mathematics & Computation. This illustrates the common pattern of the model predicting more uncited and highly cited articles but less medium cited articles than found in the empirical distribution.

There is clear evidence of a systematically incorrect shape for the hooked power law (Table 2). The model tends to substantially overestimate the number of low cited articles (see Figure 2 for an example) and underestimate the number of highly cited articles. Hence the hooked power law has the wrong shape to model the number of low cited articles. Figure 2 shows one of the few cases in which the hooked power law fits better than the discretised lognormal distribution.

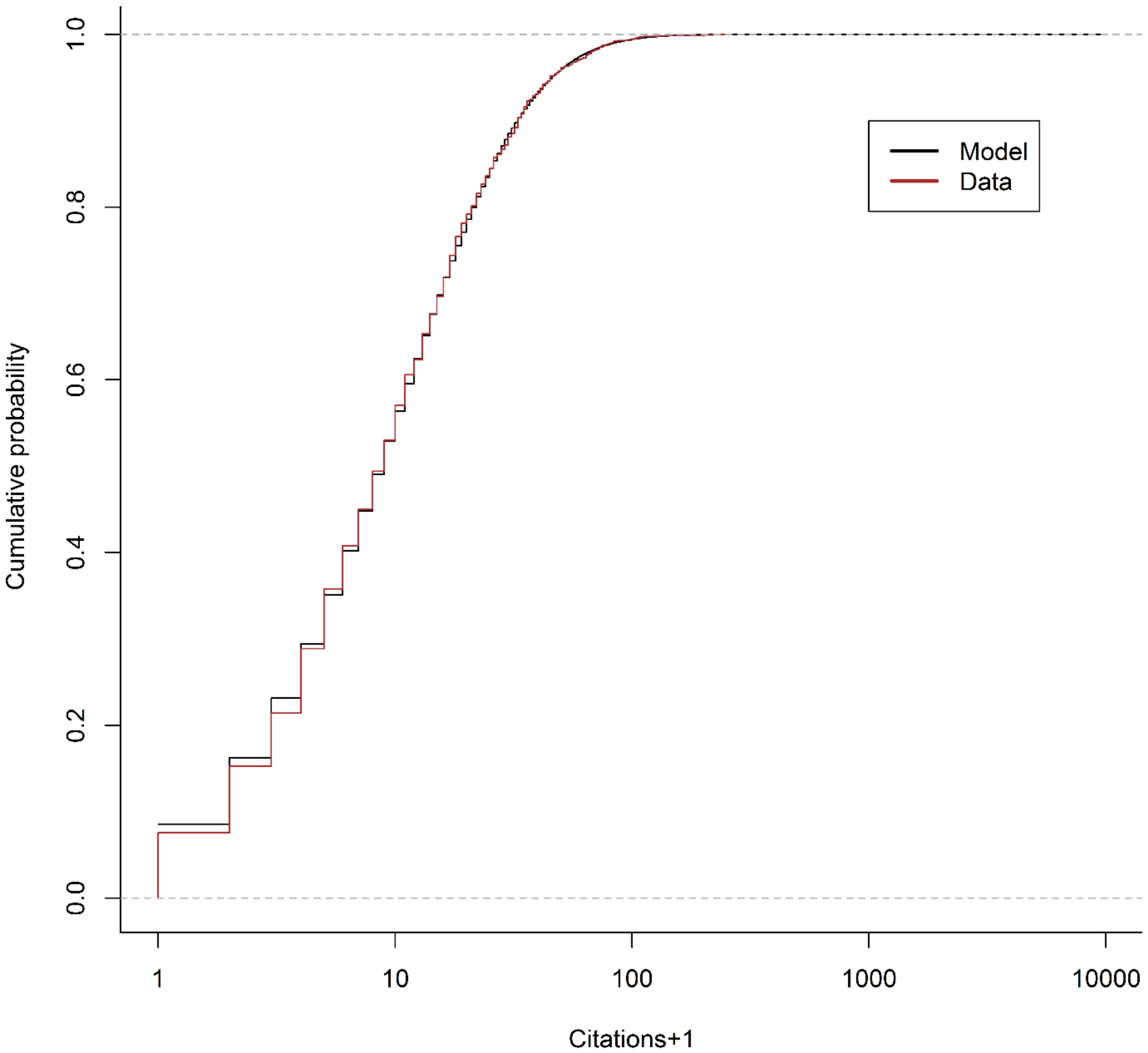

**Figure 2**. The empirical and hooked power law model cumulative distribution functions for the journal Applied Mathematics & Computation. This illustrates the common pattern of the model predicting less uncited articles than found in the empirical distribution.

**Table 2**. The largest magnitude of the empirical subtract the theoretical cumulative distribution functions for the discretised lognormal and the hooked power law. S1 – S4 are four approximately equally sized intervals on a logarithmic scale from the lowest citation count to the highest empirical distribution citation count, where S1 contains the lowest citation counts and S4 contains the highest (e.g., four equally sized x-axis intervals from 1 to about 200 in Figure 1). Positive values indicate that the theoretical distribution underestimates the number of citations within the set.

| Journal | ID* | S1 Ln | S2 Ln | S3 Ln | S4 Ln | S1 hk | S2 hk | S3 hk | S4 hk |
|---|---|---|---|---|---|---|---|---|---|
| Acta Crystallographica Section E | 48 | -1% | -1% | 0% | 0% | -3% | 1% | 1% | 0% |
| Angewandte Chemie | 43 | 1% | -2% | 2% | 1% | -6% | -11% | -5% | 3% |
| Applied Mathematics & Computation | 28 | 2% | -3% | -2% | 1% | -2% | 1% | 1% | 0% |
| Applied Physics Letters | 3 | 1% | -2% | 1% | 0% | -5% | -5% | 2% | 0% |
| Applied Surface Science | 30 | 1% | -2% | 1% | 0% | -5% | -3% | 3% | 0% |
| Astronomy & Astrophysics | 4 | 1% | -2% | 1% | 1% | -6% | -6% | 2% | -1% |
| Astrophysical Journal | 5 | 1% | -2% | 1% | 0% | -7% | -8% | 4% | -1% |
| Biochemical & Biophysical Res. Comm. | 6 | 0% | -2% | 1% | 0% | -10% | -10% | 5% | 0% |
| Biochemistry | 7 | 1% | -2% | 2% | 0% | -11% | -15% | 6% | 0% |
| Bioorganic & Medicinal Chemistry Lett. | 37 | 1% | -1% | -2% | 2% | -9% | -14% | -11% | 7% |
| Brain Research | 8 | 1% | -2% | 1% | 0% | -9% | -9% | 4% | 0% |
| Cancer Research | 9 | 0% | -1% | 2% | -1% | -7% | -14% | -9% | 2% |
| Chemical Physics Letters | 10 | 1% | -3% | 2% | 0% | -6% | -7% | 4% | -1% |
| Chinese J. of Clinical Rehabilitation | 49 | 0% | 0% | 0% | 0% | -1% | 0% | 0% | 0% |
| Geophysical Research Letters | 27 | 1% | -3% | 2% | 1% | -7% | -8% | 4% | -1% |
| Inorganic Chemistry | 11 | 0% | -2% | -2% | 1% | -8% | -11% | -7% | 3% |
| Jane's Defence Industry | 42 | 0% | 0% | 0% | 0% | 0% | 0% | 0% | 0% |
| Jane's Defence Weekly | 31 | 0% | 0% | 0% | 0% | 0% | 0% | 0% | 0% |
| Japanese J. of Applied Physics Part 1 | 12 | -2% | -2% | 1% | 0% | -2% | 2% | 0% | 0% |
| Jisuanji Gongcheng Computer Eng. | 38 | 0% | 0% | 0% | 0% | 0% | 0% | -1% | 0% |
| J.of Agricultural & Food Chemistry | 13 | 1% | -3% | -3% | 0% | -9% | -12% | -5% | 1% |
| J. of Applied Physics | 14 | 1% | -2% | 1% | 1% | -2% | 2% | 1% | 0% |
| J. of Applied Polymer Science | 15 | 1% | -3% | 2% | 1% | -5% | -5% | 2% | 0% |
| J. of Biological Chemistry | 16 | 0% | -1% | 1% | 0% | -7% | -15% | -9% | 3% |
| J. of Chemical Physics | 17 | -1% | -2% | 2% | 0% | -8% | -7% | 3% | 0% |
| J. of Immunology | 18 | 1% | -2% | 2% | 1% | -8% | -13% | 6% | 1% |
| J. of Neuroscience | 32 | 0% | -1% | 2% | -1% | -6% | -16% | -14% | 4% |
| J. of Organic Chemistry | 1 | 1% | -2% | -2% | 1% | -9% | -12% | -6% | 2% |
| J. of Physical Chemistry A | 40 | 1% | -2% | 1% | 0% | -9% | -9% | 4% | -1% |
| J. of Physical Chemistry B | 44 | 0% | -1% | 2% | 0% | -8% | -8% | 3% | -1% |
| J. of Power Sources | 33 | 1% | -4% | -4% | 2% | -5% | -6% | 4% | 2% |
| J. of the American Chemical Soc. | 2 | 0% | -1% | 1% | 0% | -8% | -11% | 5% | -1% |
| J. of Virology | 19 | 0% | 0% | 1% | -1% | -7% | -12% | -10% | 4% |
| Langmuir | 34 | 1% | -2% | 1% | 0% | -7% | -8% | 4% | -1% |
| Macromolecules | 20 | 1% | -1% | -2% | 1% | -6% | -10% | 4% | 2% |
| Materials Science & Eng. A | 36 | 1% | -2% | -2% | 1% | -5% | -5% | 3% | -1% |
| Monthly Not. R. Astronomical Soc. | 23 | 1% | -2% | 1% | 0% | -6% | -5% | 3% | -1% |
| Nuclear Instruments & Meth. Physics A | 29 | -1% | -2% | 1% | 0% | -3% | 2% | 1% | 0% |
| Optics Express | 50 | 1% | -3% | -3% | 1% | -4% | -4% | 2% | 0% |
| Organic Letters | 45 | 0% | -1% | 2% | -1% | -8% | -13% | -7% | 3% |
| Physica B Condensed Matter | 35 | -1% | 2% | 1% | 0% | 0% | 2% | 1% | 0% |
| Physical Review A | 39 | 1% | -3% | -2% | 1% | -4% | -5% | 2% | 0% |
| Physical Review B | 41 | -1% | -2% | 1% | 0% | -6% | -6% | 3% | 0% |
| Physical Review D | 47 | 1% | -4% | 1% | 1% | -2% | 2% | 2% | 0% |
| Physical Review E | 46 | -1% | -2% | 1% | 0% | -5% | -4% | 2% | 0% |
| Physical Review Letters | 22 | 1% | -2% | 1% | 0% | -8% | -8% | 4% | 0% |
| PNAS | 21 | 1% | -4% | 2% | 1% | -6% | -12% | 5% | 0% |
| Tetrahedron | 24 | 1% | -2% | 2% | 0% | -12% | -14% | 6% | 0% |
| Tetrahedron Letters | 25 | 0% | -1% | -1% | -1% | -10% | -11% | 4% | 2% |
| Thin Solid Films | 26 | 1% | -3% | -1% | 1% | -5% | -4% | 3% | 0% |
| Mean | | 0.4% | -1.8% | 0.4% | 0.3% | -5.8% | -7.0% | 0.6% | 0.6% |
| Median | | 0.6% | -1.8% | 1.1% | 0.2% | -6.2% | -7.4% | 2.3% | -0.2% |
| Total >0 | | 40 | 5 | 35 | 28 | 2 | 9 | 37 | 21 |
| Total <0 | | 10 | 45 | 15 | 22 | 48 | 41 | 13 | 29 |
| Total >1% | | 10 | 1 | 29 | 8 | 0 | 7 | 34 | 14 |
| Total <1% | | 4 | 40 | 11 | 0 | 45 | 39 | 10 | 0 |
| Total ≥-1% and ≤1% | | 36 | 9 | 10 | 42 | 5 | 4 | 6 | 36 |

* The ID column contains the number of the image file in the online supplement cited in Appendix B.

# 4. Discussion

Previous results have suggested that the hooked power law fits better than the discretised lognormal, especially for older data (Eom & Fortunato, 2011; Evans, Kaube, & Hopkins, 2012; Radicchi, Fortunato, & Castellano, 2008; Thelwall & Wilson, 2014a; Thelwall, 2016a). In this context, the dominance of the discretised lognormal distribution as the best fit for individual large journals reverses these findings and suggests that the reason for the previously-found relatively good fits of the hooked power law for whole categories is that subject categories that include multiple journals tend to mix different distributions. This may occur through the inclusion of journals with different specialisms, albeit within a single subject category (although journals also span multiple Scopus categories). It may also occur through the inclusion of journals with different languages, because more nationally-focused journals may have different citation cultures and may also have a low proportion of their citations indexed in the Web of Science and Scopus.

The field purity of the sets of articles from the 50 journals studied here is only relative. As shown in Table 3 in Appendix B, few of the journals seem to represent narrow specialisms. Even those classified in this table as narrow because they are only in one Scopus category, such as Physical Review A: Atomic, Molecular and Optical Physics, may well cover multiple specialisms within their field. In the case of this journal, it seems to cover three specialisms within Physics. Hence, the discussion in the paragraph above should carry the caveat that large journals are a source of *relatively* pure citation count data. Nevertheless, the trend (e.g., in comparison to: Thelwall, 2016a) seems to be that the purer the set of articles, the more likely that the discretised lognormal fits better than the shifted power law. This contradicts a previous analysis of American Physical Society journals (Eom & Fortunato, 2011), which found the opposite. The reason for this may be that the citation counts for these journals were derived from the American Physical Society database rather than a multidisciplinary source, such as Scopus because four of these ten journals are in the set analysed in the current paper (Physical Review A,B,D,E).

Even if all journals within a subject category exactly followed the same one of the two distributions, but with different parameters, then it does not follow that the distribution of the entire subject category would follow the same distribution. For example the sum of two lognormal distributions could be bimodal rather than unimodal. This property is inherited from the same property of the normal distribution. See the see the online supplement cited in Appendix B for evidence that the sum of two hooked power laws is not necessarily another hooked power law, based upon proving this for the simpler continuous case: the Lomax distribution. Nevertheless, it seems likely that if the distributions for individual journals are not too dissimilar then a similar distribution should fit the entire subject category well. More investigations are therefore needed to assess how the distributions for individual journals relate to those for entire subject categories.

The results are limited by the analysis being restricted to large journals and it is possible that these have characteristics inherent in their size that affect the results, although this seems unlikely. The results are also limited in the range of different subjects because there are no representatives of arts, humanities and social sciences. It seems plausible that these areas could have citation cultures that do not fit the discretised lognormal as well. Given the scarcity of large journals in these areas, it may be difficult to test this hypothesis with pure data sets.

## 5. Conclusions

The primacy of the discretised lognormal distribution for relatively pure citation distributions is a useful result for those wishing to conduct theoretical analyses of the properties of citation-based indicators. This is because the continuous lognormal distribution is a well-understood and has tractable properties that are inherited from the underlying normal distribution. Even though a discretised variant is the one that fits citation counts, it seems reasonable to use the continuous distribution approximation in order to be able to mathematically analyse citation indicators. In contrast, although the Lomax distribution underlying the hooked power law has also been studied in the past, it has instabilities (such as the realistic possibility of very large optimal $\alpha$ values) that make it more difficult to work with. It is also seems to be less tractable analytically for mathematical analyses.

This paper also introduces, and makes available online (see Appendix B), a new more powerful method for fitting the hooked power law to citation-like data sets. This allows hooked power law distributions with very large $\alpha$ values to be fitted, although very slowly. For example, it may take over a week to fit a single distribution with $\alpha$=10000. This may nevertheless be useful for analyses of the citation counts of sets of articles from individual fields, where the hooked power law tends to fit better than he discretised lognormal distribution.

## 6. References

Akaike, H. (1974). A new look at the statistical model identification. IEEE Transactions on Automatic Control, 19(6), 716-723.

Brzezinski, M. (2015). Power laws in citation distributions: Evidence from Scopus. Scientometrics, 103(1), 213-228.

Burnham, K. P., & Anderson, D. R. (2002). Model selection and multimodel inference: a practical-theoretic approach. New York, NY: Springer.

Burrell, Q. L. (2008). Extending Lotkaian informetrics. Information Processing & Management, 44(5), 1794-1807.

Clauset, A., Shalizi, C. R., & Newman, M. E. (2009). Power-law distributions in empirical data. SIAM review, 51(4), 661-703.

Didegah, F., & Thelwall, M. (2013). Which factors help authors produce the highest impact research? Collaboration, journal and document properties. Journal of Informetrics, 7(4), 861-873.

Eom, Y. H., & Fortunato, S. (2011). Characterizing and modeling citation dynamics. PLoS One, 6(9), e24926.

Evans, T. S., Kaube, B. S., & Hopkins, N. (2012). Universality of performance indicators based on citation and reference counts. Scientometrics, 93(2), 473-495.

Fousse, L., Hanrot, G., Lefèvre, V., Pélissier, P., & Zimmermann, P. (2007). MPFR: A multiple-precision binary floating-point library with correct rounding. ACM Transactions on Mathematical Software, 33(2), article 13. doi:10.1145/1236463.1236468.

Garfield, E. (2006). The history and meaning of the journal impact factor. Jama, 295(1), 90-93.

Gazni, A. (2011). Are the abstracts of high impact articles more readable? Investigating the evidence from top research institutions in the world. Journal of Information Science, 37 (3), 273-281.

Gillespie, C.S. (2015). Fitting heavy tailed distributions: the poweRlaw package. Journal of Statistical Software, 64(2), 1-16. http://www.jstatsoft.org/v64/i02/paper

Glänzel, W. (2001). National characteristics in international scientific co-authorship relations. Scientometrics, 51(1), 69-115.

Hirsch, J. E. (2005). An index to quantify an individual's scientific research output. Proceedings of the National academy of Sciences of the United States of America, 102(46), 16569-16572.

IEEE (2008). 754-2008 - IEEE standard for floating-point arithmetic. http://standards.ieee.org/findstds/standard/754-2008.html

Li, J., Burnham, J.F., Lemley, T., & Britton, R.M. (2010). Citation analysis: Comparison of Web of Science, Scopus, SciFinder, and Google Scholar. Journal of Electronic Resources in Medical Libraries, 7(3), 196-217.

Limpert, E., Stahel, W.A. & Abbt, M. (2001). Lognormal distribution across sciences: Key and clues. Bioscience, 51(5), 341-351.

Lomax, K. S. (1954). Business failures: Another example of the analysis of failure data. Journal of the American Statistical Association, 49(268), 847–852.

López-Illescas, C., de Moya-Anegón, F., & Moed, H. F. (2008). Coverage and citation impact of oncological journals in the Web of Science and Scopus. Journal of Informetrics, 2(4), 304-316.

Low, W. J., Thelwall, M. & Wilson, P. (2015). Stopped sum models for citation data. In Salah, A.A., Y. Tonta, A.A. Akdag Salah, C. Sugimoto, U. Al (Eds.), Proceedings of ISSI 2015 Istanbul: 15th International Society of Scientometrics and Informetrics Conference. Istanbul, Turkey: Bogaziçi University Printhouse (pp. 184-194).

Machler, M. (2015). Arbitrarily accurate computation with R: The Rmpfr package. https://cran.r-project.org/web/packages/Rmpfr/vignettes/Rmpfr-pkg.pdf

Merton, R. K. (1968). The Matthew effect in science. Science, 159(3810), 56-63.

Merton, R. K. (1973). The sociology of science: Theoretical and empirical investigations. Chicago, IL: University of Chicago press.

Moed, H. F., & Visser, M. S. (2008). Appraisal of citation data sources. Centre for Science and Technology Studies, Leiden University. Retrieved November 29, 2014, from http://www.hefce.ac.uk/media/hefce/content/pubs/indirreports/2008/missing/Appraisal%20of%20Citation%20Data%20Sources.pdf.

Onodera, N., & Yoshikane, F. (2015). Factors affecting citation rates of research articles. Journal of the Association for Information Science and Technology, 66(4), 739-764.

Pennock, D. M., Flake, G. W., Lawrence, S., Glover, E. J., & Giles, C. L. (2002). Winners don't take all: Characterizing the competition for links on the web. Proceedings of the national academy of sciences, 99(8), 5207-5211.

Radicchi, F., Fortunato, S., & Castellano, C. (2008). Universality of citation distributions: Toward an objective measure of scientific impact. Proceedings of the National Academy of Sciences, 105(45), 17268-17272.

Redner, S. (1998). How popular is your paper? An empirical study of the citation distribution. The European Physical Journal B-Condensed Matter and Complex Systems, 4(2), 131-134.

Revol, N., & Rouillier, F. (2005). Motivations for an arbitrary precision interval arithmetic and the MPFI library. Reliable Computing, 11(4), 275-290.

Thelwall, M. & Fairclough, R. (2015). Geometric journal impact factors correcting for individual highly cited articles. Journal of Informetrics, 9(2), 263–272.

Thelwall, M. & Wilson, P. (2014a). Distributions for cited articles from individual subjects and years. Journal of Informetrics, 8(4), 824-839.

Thelwall, M. & Wilson, P. (2014b). Regression for citation data: An evaluation of different methods. Journal of Informetrics, 8(4), 963–971.

Thelwall, M. (2016a). The discretised lognormal and hooked power law distributions for complete citation data: Best options for modelling and regression. Journal of Informetrics, 10(2), 336-346. doi:10.1016/j.joi.2015.12.007

Thelwall, M. (2016b). Are the discretised lognormal and hooked power law distributions plausible for citation data? Journal of Informetrics, 10(2), 454-470. doi:10.1016/j.joi.2016.03.001

Vuong, Q. H. (1989). Likelihood ratio tests for model selection and non-nested hypotheses. Econometrica: Journal of the Econometric Society, 57(2), 307-333.

Waltman, L., van Eck, N. J., van Leeuwen, T. N., Visser, M. S., & van Raan, A. F. (2011). Towards a new crown indicator: Some theoretical considerations. Journal of Informetrics, 5(1), 37-47.

Wang, D., Song, C., & Barabási, A. L. (2013). Quantifying long-term scientific impact. Science, 342(6154), 127-132.

Wu, Y., Fu, T. Z., & Chiu, D. M. (2014). Generalized preferential attachment considering aging. Journal of Informetrics, 8(3), 650-658.

Yao, Z., Peng, X. L., Zhang, L. J., & Xu, X. J. (2014). Modeling nonuniversal citation distributions: the role of scientific journals. Journal of Statistical Mechanics: Theory and Experiment, 2014(4), P04029.

Zitt, M. (2012). The journal impact factor: Angel, devil, or scapegoat? A comment on JK Vanclay's article 2011. Scientometrics, 92(2), 485-503.

## Appendix A: Accuracy considerations

The parameters of the hooked power law are problematic to calculate in practice when they are large. This is because the numbers in the calculations required to obtain the parameters are so small when the parameters are large that rounding errors can occur, leading to incorrect parameter estimates.

The standard format for representing (floating point) numbers in computers is the binary62 IEEE 754 standard (IEEE, 2008). The smallest number that can be stored without reducing the number of significant digits available following this standard is $1 \times 10^{-308}$. This can be checked for the statistical software R on any computer through the *double.xmin* value reported by the command *noquote(unlist(format(.Machine)))*. Smaller numbers can be stored at reduced accuracy (loosing significant digits and effectively truncating the mantissa by allowing them to have leading zeros) but by $1 \times 10^{-324}$ all numbers are truncated to zero. Thus in R the command x=10^-324 is equivalent to x=0. The hooked power law distribution requires an approximation to $\sum_{n=1}^{\infty}(B+n)^{-\alpha}$, such as with $\sum_{n=1}^{10000}(B+n)^{-\alpha}$. The accuracy of this calculation is compromised if any of the values must be calculated at reduced accuracy. The smallest number in this sum is $(B+10000)^{-\alpha}$ and so this should be at least $1 \times 10^{-308}$. Using base ten logarithms, $(B+10000)^{-\alpha} > 1 \times 10^{-308}$ is equivalent to $\log_{10}(B+10000)^{-\alpha} > \log_{10}(1 \times 10^{-308})$, which is equivalent to $-\alpha\log_{10}(B+10000) > -308$. Assuming that B is small relative to 10,000 then $\log_{10}(B+10000) \approx 4$ and so the key property is $\alpha \times -4 > -308$ or $\alpha > \frac{308}{4} = 77$. Hence, as α approaches 77, the accuracy of the hooked power law model starts to fall. Similar calculations show that as α and B become large (e.g., α=100 and B=200), the sum $\sum_{n=1}^{10000}(B+n)^{-\alpha}$ will be equal to zero due to rounding errors, making it impossible to fit the model to the data. Hence a method is needed to allow higher precision calculations in order to fit the hooked power law when its optimal parameters are large.

Increased accuracy is available from extended precision arithmetic with special purpose software but, because extended precision is not supported by most current processors, it is about a thousand times slower (author's tests in R). A standard extended precision software library is the Multiple Precision Floating-point Reliable (MPFR) C library (Fousse, Hanrot, Lefèvre, Pélissier, & Zimmermann, 2007; Revol & Rouillier, 2005) from www.mpfr.org, which can be imported into R via the Rmpfr package (Machler, 2015). This is built into the R code supplied with the current paper. To maximise speed, the R code supplied automatically switches to higher precision calculations when necessary but uses standard precision otherwise. On a few of the data sets the model fitting and associated other tests did not finish after a month of continuous running and so it seems to be too slow for use in practice on some data sets. Thus, the hooked power law is impractical for some data sets with current standard computing technology. Although it does not seem to be explicitly documented, the largest number that can be represented in 128 bit arithmetic with MPFR is at least $10^{320,000,000}$ (this can be checked with the R code: *q = mpfr(10,128); q = q^320000000 * pi; q;)*. To enable results to be calculated in a reasonable amount of time, a limit of 10,000 was set for the alpha value and the fitting algorithm terminated once this value had been reached.

## Appendix B: Additional data

An online supplement with graphs for all 100 models, the spreadsheet with the results, the raw data and the R code used is available online at: https://figshare.com/articles/Citation_count_distributions_for_large_monodisciplinary_journals/3479129 doi:10.6084/m9.figshare.3479129

**Table 3**. The 50 journals with the most journal articles indexed in 2006 in Scopus. [Type: M=multidisciplinary; n=number of fields; B=Broad field; N=narrow field]. The Scopus brief subject area column contains the classification shown in search results pages for matching articles within the journal. The Scopus full subject area column contains the classification in the journal information page in Scopus.

| Journal | Articles | Type* | Scopus brief subject area | Scopus full subject area |
|---|---|---|---|---|
| Applied Physics Letters | 6,103 | B | Physics & Astronomy | Physics & Astronomy: Physics & Astronomy (miscellaneous) |
| Physical Review B Condensed Matter and Materials Physics | 5,603 | 2 | Physics & Astronomy | Materials Science: Electronic, Optical & Magnetic Materials; Physics & Astronomy: Condensed Matter Physics |
| Journal of Biological Chemistry | 4,307 | B | Biochemistry, Genetics & Molecular Biology | Biochemistry, Genetics & Molecular Biology: Biochemistry; Biochemistry, Genetics & Molecular Biology: Cell Biology; Biochemistry, Genetics & Molecular Biology: Molecular Biology |
| Acta Crystallographica Section E Structure Reports Online | 4,243 | 3 | Biochemistry, Genetics & Molecular Biology; Physics & Astronomy | Chemistry; Materials Science; Physics & Astronomy: Condensed Matter Physics |
| Physical Review Letters | 3,759 | 2 | Physics & Astronomy | Medicine; Physics & Astronomy |
| Journal of Physical Chemistry B | 3,617 | 2 | Chemistry | Chemistry: Physical & Theoretical Chemistry; Materials Science: Materials Chemistry; Materials Science: Surfaces, Coatings & Films; Medicine |
| Journal of Applied Physics | 3,595 | B | Physics & Astronomy | Physics & Astronomy |
| Proceedings of the National Academy of Sciences of the United States of America | 3,297 | M | Multidisciplinary; Biochemistry, Genetics & Molecular Biology | Multidisciplinary |
| Journal of the American Chemical Society | 3,256 | 2 | Chemistry | Biochemistry, Genetics & Molecular Biology: Biochemistry; Chemical Engineering: Catalysis; Chemical Engineering: Colloid & Surface Chemistry; Chemistry; Medicine |
| Journal of Chemical Physics | 2,870 | 3 | Physics & Astronomy | Chemistry: Physical & Theoretical Chemistry; Medicine; |
| | | | | Physics & Astronomy |
| Astrophysical Journal | 2,690 | 2 | Earth & Planetary Sciences;<br>Physics & Astronomy | Earth & Planetary Sciences: Space & Planetary Science;<br>Physics & Astronomy: Astronomy & Astrophysics |
| Chinese Journal of Clinical Rehabilitation | 2,671 | N | Medicine | Medicine: Rehabilitation |
| Physical Review E Statistical Nonlinear and Soft Matter Physics | 2,451 | 2 | Mathematics; Physics & Astronomy | Mathematics: Statistics & Probability;<br>Physics & Astronomy: Condensed Matter Physics;<br>Physics & Astronomy: Statistical & Nonlinear Physics |
| Journal of Applied Polymer Science | 2,439 | 2 | Materials Science | Chemistry; Materials Science: Materials Chemistry;<br>Materials Science: Polymers & Plastics;<br>Materials Science: Surfaces, Coatings & Films |
| Biochemical and Biophysical Research Communications | 2,363 | 2 | Biochemistry, Genetics & Molecular Biology | Biochemistry, Genetics & Molecular Biology: Biochemistry;<br>Biochemistry, Genetics & Molecular Biology: Biophysics;<br>Biochemistry, Genetics & Molecular Biology: Cell Biology;<br>Biochemistry, Genetics & Molecular Biology: Molecular Biology;<br>Medicine |
| Physical Review D Particles Fields Gravitation and Cosmology | 2,305 | 2 | Mathematics;<br>Physics & Astronomy | Physics & Astronomy: Nuclear & High Energy Physics |
| Physical Review A Atomic Molecular and Optical Physics | 2,080 | N | Physics & Astronomy | Physics & Astronomy: Atomic & Molecular Physics, & Optics |
| Tetrahedron Letters | 1,988 | 3 | Biochemistry, Genetics & Molecular Biology;<br>Chemistry;<br>Pharmacology, Toxicology & Pharmaceutics | Biochemistry, Genetics & Molecular Biology: Biochemistry;<br>Chemistry: Organic Chemistry;<br>Pharmacology, Toxicology & Pharmaceutics: Drug Discovery |
| Jane's Defence Weekly | 1,976 | 2 | Engineering | Business, Management & Accounting: Strategy & Management;<br>Engineering: Aerospace Engineering;<br>Engineering: Automotive Engineering;<br>Engineering: Engineering (miscellaneous);<br>Engineering: Mechanical Engineering |
| Jisuanji Gongcheng Computer Engineering | 1,947 | B | Computer Science | Computer Science: Computational Theory & Mathematics;<br>Computer Science: Computer Graphics & CAD;<br>Computer Science: Computer Networks & Communications;<br>Computer Science: Hardware & Architecture;<br>Computer Science: Software |
| Astronomy and Astrophysics | 1,865 | 2 | Earth & Planetary Sciences | Earth & Planetary Sciences: Space & Planetary Science; |
| | | | | Physics & Astronomy: Astronomy & Astrophysics |
| Journal of Immunology | 1,822 | 2 | Immunology & Microbiology | Immunology & Microbiology: Immunology;<br>Medicine |
| Japanese Journal of Applied Physics Part 1 Regular Papers and Short Notes and Review Papers | 1,819 | 2 | Physics & Astronomy | Engineering;<br>Physics & Astronomy |
| Journal of Physical Chemistry A | 1,715 | 2 | Chemistry | Chemistry: Physical & Theoretical Chemistry;<br>Medicine |
| Langmuir | 1,696 | 2 | Chemical Engineering; Chemistry | Chemistry: Electrochemistry;<br>Chemistry: Spectroscopy; Materials Science; Medicine;<br>Physics & Astronomy: Condensed Matter Physics;<br>Physics & Astronomy: Surfaces & Interfaces |
| Chemical Physics Letters | 1,651 | 2 | Chemistry; Physics & Astronomy | Chemistry: Physical & Theoretical Chemistry;<br>Physics & Astronomy |
| Geophysical Research Letters | 1,637 | N | Earth & Planetary Sciences | Earth & Planetary Sciences;<br>Earth & Planetary Sciences: Geophysics |
| Biochemistry | 1,600 | 2 | Biochemistry, Genetics & Molecular Biology | Biochemistry, Genetics & Molecular Biology: Biochemistry;<br>Medicine |
| Nuclear Instruments and Methods in Physics Research Section A Accelerators Spectrometers Detectors and Associated Equipment | 1,572 | N | Physics & Astronomy | Physics & Astronomy: Instrumentation;<br>Physics & Astronomy: Nuclear & High Energy Physics |
| Applied Surface Science | 1,545 | 2 | Chemistry;<br>Materials Science; Physics & Astronomy | Materials Science: Surfaces, Coatings & Films |
| Organic Letters | 1,524 | 3 | Biochemistry, Genetics & Molecular Biology | Biochemistry, Genetics & Molecular Biology: Biochemistry;<br>Chemistry: Organic Chemistry;<br>Chemistry: Physical & Theoretical Chemistry;<br>Medicine |
| Materials Science and Engineering A | 1,490 | 3 | Materials Science | Engineering: Mechanical Engineering;<br>Engineering: Mechanics of Materials;<br>Materials Science;<br>Physics & Astronomy: Condensed Matter Physics |
| Journal of Power Sources | 1,476 | 3 | Energy;<br>Chemistry;<br>Materials Science; Engineering | Chemistry: Physical & Theoretical Chemistry;<br>Energy: Energy Engineering & Power Technology;<br>Energy: Renewable Energy, Sustainability & the Environment;<br>Engineering: Electrical & Electronic Engineering |
| Journal of Organic Chemistry | 1,472 | B | Chemistry | Chemistry: Organic Chemistry |
| Journal of Agricultural and Food Chemistry | 1,449 | 3 | Agricultural & Biological Sciences;<br>Chemistry | Agricultural & Biological Sciences;<br>Chemistry;<br>Medicine |
| Inorganic Chemistry | 1,432 | 2 | Chemistry | Chemistry: Inorganic Chemistry;<br>Chemistry: Physical & Theoretical Chemistry;<br>Medicine |
| Cancer Research | 1,428 | 2 | Biochemistry, Genetics & Molecular Biology; Medicine | Biochemistry, Genetics & Molecular Biology: Cancer Research;<br>Medicine;<br>Medicine: Oncology |
| Brain Research | 1,375 | 3 | Biochemistry, Genetics & Molecular Biology; Medicine; Neuroscience | Biochemistry, Genetics & Molecular Biology: Developmental Biology;<br>Biochemistry, Genetics & Molecular Biology: Molecular Biology;<br>Medicine;<br>Medicine: Neurology (clinical);<br>Neuroscience |
| Angewandte Chemie International Edition | 1,361 | 3 | Chemistry | Chemical Engineering: Catalysis;<br>Chemistry;<br>Medicine |
| Monthly Notices of the Royal Astronomical Society | 1,354 | 2 | Earth & Planetary Sciences | Earth & Planetary Sciences: Space & Planetary Science;<br>Physics & Astronomy: Astronomy & Astrophysics |
| Physica B Condensed Matter | 1,351 | 3 | Materials Science; Physics & Astronomy | Engineering: Electrical & Electronic Engineering;<br>Materials Science: Electronic, Optical & Magnetic Materials;<br>Physics & Astronomy: Condensed Matter Physics |
| Optics Express | 1,349 | N | Physics & Astronomy | Physics & Astronomy: Atomic & Molecular Physics, & Optics |
| Journal of Neuroscience | 1,326 | 2 | Medicine;<br>Neuroscience | Medicine;<br>Neuroscience |
| Jane's Defence Industry | 1,320 | N | Engineering | Engineering: Engineering (miscellaneous) |
| Thin Solid Films | 1,279 | 2 | Materials Science; Physics & Astronomy | Materials Science: Electronic, Optical & Magnetic Materials;<br>Materials Science: Materials Chemistry;<br>Materials Science: Metals & Alloys;<br>Materials Science: Surfaces, Coatings & Films;<br>Physics & Astronomy: Surfaces & Interfaces |
| Tetrahedron | 1,275 | 3 | Biochemistry, Genetics & Molecular Biology; Chemistry; Pharmacology, Toxicology & | Biochemistry, Genetics & Molecular Biology: Biochemistry;<br>Chemistry: Organic Chemistry;<br>Pharmacology, Toxicology & Pharmaceutics: Drug Discovery |
| | | | Pharmaceutics | |
| Macromolecules | 1,264 | 2 | Chemistry; Materials Science | Chemistry: Inorganic Chemistry; Chemistry: Organic Chemistry; Materials Science: Materials Chemistry; Materials Science: Polymers & Plastics |
| Bioorganic and Medicinal Chemistry Letters | 1,264 | 4 | Biochemistry, Genetics & Molecular Biology; Chemistry; Pharmacology, Toxicology & Pharmaceutics | Biochemistry, Genetics & Molecular Biology: Biochemistry; Biochemistry, Genetics & Molecular Biology: Clinical Biochemistry; Biochemistry, Genetics & Molecular Biology: Molecular Biology; Biochemistry, Genetics & Molecular Biology: Molecular Medicine; Chemistry: Organic Chemistry; Medicine; Pharmacology, Toxicology & Pharmaceutics: Drug Discovery; Pharmacology, Toxicology & Pharmaceutics: Pharmaceutical Science |
| Journal of Virology | 1,263 | 2 | Immunology & Microbiology | Immunology & Microbiology: Immunology; Immunology & Microbiology: Virology; Medicine |
| Applied Mathematics and Computation | 1,249 | B | Mathematics | Mathematics: Applied Mathematics; Mathematics: Computational Mathematics |